\documentclass[twocolumn,showpacs,prl]{revtex4}
\usepackage{graphicx}
\usepackage{latexsym}

\begin{document}

\title{Magnetic Field Induced Spin Polarization of AlAs Two-dimensional Electrons}

\author{E.\ P.\ De Poortere, E.\ Tutuc, Y.\ P.\ Shkolnikov, K.\ Vakili, and M.\ Shayegan}
\address{Department of Electrical Engineering, Princeton University,
Princeton, New Jersey 08544}

\date{\today}

\begin{abstract}

Two-dimensional (2D) electrons in an in-plane magnetic field become fully spin polarized above a field $B_P$,
which we can determine from the in-plane magnetoresistance. We perform such measurements in modulation-doped AlAs
electron systems, and find that the field $B_P$ increases approximately linearly with 2D electron density. These
results imply that the product $|g^*|m^*$, where $g^*$ is the effective $g$-factor and $m^*$ the effective mass,
is a constant essentially independent of density. While the deduced $|g^*|m^*$ is enhanced relative to its band
value by a factor of $\sim$ 4, we see no indication of its divergence as 2D density approaches zero. These
observations are at odds with results obtained in Si-MOSFETs, but qualitatively confirm spin polarization studies
of 2D GaAs carriers.

\end{abstract}


\maketitle

The ground state properties of dilute 2D carrier systems are still under close scrutiny, and several issues, such
as the existence of a metal-insulator transition and the occurrence of a ferromagnetic instability at zero
magnetic fields, remain largely unresolved. At the lowest densities ($n$), calculations predict that 2D electrons
settle into a Wigner crystal \cite{tanatar89}, while at the highest $n$, a paramagnetic Fermi liquid model most
likely describes the ground state of the 2D electron system (2DES). At intermediate densities, a Stoner transition
to a fully spin polarized ground state may take place \cite{bloch29,stoner38,tanatar89,varsano01,benenti01}, but
experimental evidence for such a phase has been scant so far. A semi-direct method used to probe this
ferromagnetic state involves measuring the product $|g^*|m^*$ in the 2D electron system at various densities. A
diverging $|g^*|m^*$ as $n$ approaches a critical density, would then imply that the system spontaneously
polarizes at lower $n$. Accordingly, several groups have recently studied the density dependence of $|g^*|m^*$ in
dilute 2D carrier systems, but have not yet reached an agreement. Some of these groups have reported an increasing
$|g^*|m^*$ as $n$ decreases in Si MOSFETs \cite{fang68,okamoto99,pudalov01,shashkin01,vitkalov01}. Furthermore,
scaling arguments based on in-plane magnetoresistance of electrons in the same system \cite{shashkin01,vitkalov01}
suggest that $|g^*|m^*$ diverges at a finite density, hinting at a possible Stoner instability. On the other hand,
measurements of $|g^*|m^*$ in 2D carrier systems in GaAs \cite{tutuc02,papadakis00,yoon00} point to the contrary.
Direct measurements of the magnetization of Si 2D electrons also offer no evidence for a ferromagnetic ground
state \cite{prus02}.

In an attempt to resolve these discrepancies, we have performed experiments in a new system, 2D electrons in AlAs,
in which the effect of disorder is very much reduced: we have recently obtained AlAs samples with electron
mobilities as high as 30 m$^2$/Vs \cite{depoortere02}. For similar $r_s$, defined as the ratio of average
interparticle separation to the effective Bohr radius, mobilities in AlAs are roughly an order of magnitude larger
than those in Si, and a factor 2 lower than in GaAs. We find that $|g^*|m^*$ in AlAs 2D electrons is approximately
independent of density for the measured range $0.5 < n < 5.9 \times 10^{11}$ cm$^{-2}$. Our results imply that the
ground state of the AlAs 2DES remains unpolarized at all measured densities, a conclusion in agreement with
transport experiments in GaAs and with magnetization measurements in Si \cite{prus02}.

The electronic band structure of AlAs is reminiscent of that of Si: conduction electrons in AlAs, located at the
X-point of the Brillouin zone, have an isotropic effective mass ($m_l$ = 1.0, $m_t$ = 0.2) almost identical to
that of Si, and a band $g$-factor equal to 2, as in Si \cite{vankesteren89}. In 2D however, properties of Si and
AlAs electrons are different: AlAs (100) 2D electrons in quantum wells wider than $\sim$ 45 $\AA$
\cite{yamada94,vandestadt96} occupy X-point valleys with principal axes parallel to the plane of the 2DES, so that
their cyclotron effective mass is $m_b^* = (m_l.m_t)^{1/2} =$ 0.46 \cite{lay93}, while in Si (100), out-of-plane
valleys are occupied, and $m_b^* = m_t =$ 0.2 (effective masses are given in units of the free electron mass).
Strain induced by lattice-mismatch between the AlAs quantum well and AlGaAs barriers is responsible for increasing
the energy of the out-of-plane valleys in AlAs relative the energy of the in-plane valleys, so that for wide AlAs
quantum wells, such as those used in our study, in-plane valleys are occupied. In Si MOSFETs and in narrow AlAs
quantum wells, on the other hand, out-of-plane plane valleys are populated due to the higher effective mass in the
growth direction.

Results for three samples (A-C) are presented here. These were grown by molecular beam epitaxy atop a GaAs (411)B
substrate for sample A (M393IV$\beta$-D2), and GaAs (100) substrates for samples B (M420-D3) and C (M415-J2). The
layer structures for these samples are composed of a 150 \AA\ (samples A and C) or a 120 \AA -wide (sample B) AlAs
quantum well, sandwiched between Al$_{x}$Ga$_{1-x}$As barriers, with 0.25 $<$ x $<$ 0.4, and modulation doped with
Si. We patterned all samples in a Hall bar geometry for transport measurements in magnetic fields up to 18 T,
which were conducted in a $^3$He cryostat with a base temperature of $T = 300$ mK, and in a dilution refrigerator
for $T$ down to 30 mK. Samples were mounted on a single-axis rotator stage, which allowed us to tune the angle
between the magnetic field direction and the plane of the 2D electron system. After illumination, carrier density
was varied with the help of both front- and back-gates, and was determined from Shubnikov-de Haas oscillations
measured in a perpendicular magnetic field. In sample A, electron mobility at $T = 30$ mK ranged from $\mu = 5.6$
m$^2$/Vs at $n = 1.2 \times 10^{11}$ cm$^{-2}$ to $\mu = 14$ m$^2$/Vs at $n = 5.9 \times 10^{11}$ cm$^{-2}$.

\begin{figure}
\includegraphics[scale=.42]{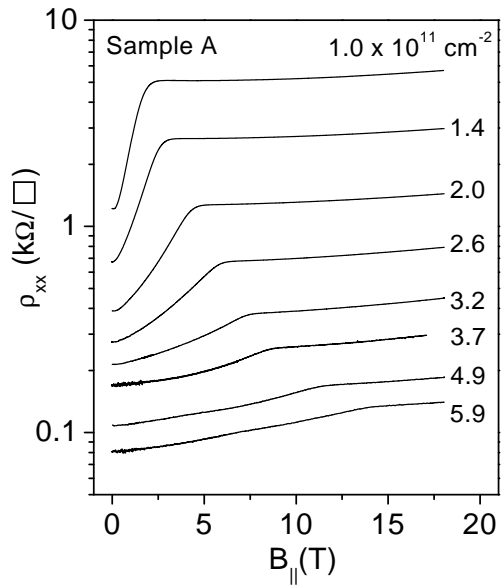}
\caption{In-plane magnetoresistance of AlAs 2D electrons for different carrier densities in sample A at $T \simeq
30$ mK. Magnetoresistance increases strongly at low fields, and saturates or increases more slowly at higher
fields, when the 2DES is fully polarized.}
\end{figure}

In-plane magnetoresistance data ($\rho_{xx}$) for sample A are plotted in Fig.\ 1 for several carrier densities.
As in Si MOSFETs, the resistivity shows a strong increase with $B_{||}$ at low fields, while it rises much more
slowly beyond a crossover field $B_P$. Based on experimental \cite{tutuc02,papadakis00,okamoto99,vitkalov00} and
theoretical \cite{dolgopolov00} work, it is believed that at $B_P$ the 2DES undergoes a transition from a
partially spin-polarized state to a fully spin-polarized state. For one of the traces shown ($n = 2.6 \times
10^{11}$ cm$^{-2}$), using a method reported previously \cite{tutuc02}, we verified that above $B_P$ the system
indeed becomes spin-polarized. We note that Pudalov {\it et al.} \cite{pudalov02} have reported that the crossover
field in Si MOSFETs is sample dependent, and conclude that $B_P$ does not reflect complete spin polarization of
the 2DES. As our results, in contrast, are reproduced in several samples, we follow the assumption that $B_P$ does
reflect full spin polarization of the 2DES. At $B = B_P$, the Fermi energy of the system is equal to its Zeeman
energy, which gives the following relationship between $B_P$ and $|g^*|m^*$:
\begin{equation}
B_P = {n \over |g^*|m^*} {2h \over e},
\end{equation} \noindent
valid for a single-valley 2DES \cite{valleyoccupation}. From this equation, and knowing $B_P$, we can thus extract
the product $|g^*|m^*$.

\begin{figure}
\includegraphics[scale=.35]{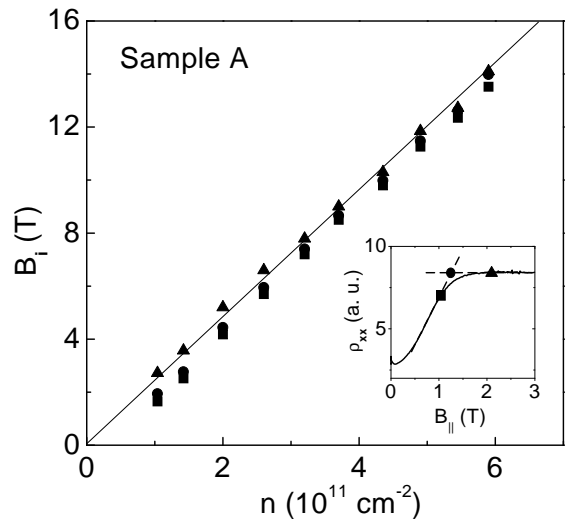}
\caption{$B_{min}$, $B_{mid}$, $B_{max}$ fields vs.\ $n$ for sample A ($T \simeq$ 30 mK). Inset: Schematic
definition of $B_i$ fields, indicated by squares, circles and triangles respectively. These fields define a range
within which spin subband depopulation likely takes place.}
\end{figure}

The crossover in magnetoresistance in Fig.\ 1 is not perfectly sharp, but can be described empirically by fitting
straight lines to the low- and high-field ranges of the trace (in a linear plot), and defining two respective
fields, $B_{min}$ and $B_{max}$, at which $\rho_{xx}$ deviates from these lines (see Fig.\ 2 inset). A deviation
ratio of 1 $\%$ is used for $B_{min}$, while we use 0.1 $\%$ for $B_{max}$ \cite{deviation}. A smaller deviation
criterion is used for $B_{max}$ than for $B_{min}$, because $\rho_{xx}$ departs more abruptly from the straight
line fit around $B_{max}$. The two deviation points are marked respectively by squares and triangles in the inset
of Fig.\ 2. A third field ($B_{mid}$), marked by a circle in Fig.\ 2, is defined as the intersection between the
two lines. For AlAs 2D electrons, full spin polarization likely takes place within the range [$B_{min}$,
$B_{max}$], though we do not know the exact relationship between the actual polarization field $B_P$ and the range
limits. In this paper we assume, as is evidenced in GaAs 2D electrons \cite{tutuc02} and holes \cite{tutuc01},
that $B_P$ is the upper field $B_{max}$. For completeness, we also provide here the range of values, [$B_{min}$,
$B_{max}$], within which $B_P$ is likely to lie. We note that our conclusion, namely, that $|g^*|m^*$ does not
increase as $n$ decreases, does not depend on the choice of $B_P$ within the range [$B_{min}$, $B_{max}$].

We plot in Fig.\ 2 the three ``$B_{i}$'' fields $B_{min}$, $B_{mid}$, and $B_{max}$, obtained from the traces in
Fig.\ 1, as a function of sample density. These values increase remarkably linearly with density, and a fit
through the upper fields $B_{max}$ appears to intercept the horizontal axis at a density small enough as to be
identified with zero, within error margins. In Fig.\ 3, similar data is plotted for sample B at $T = 300$ mK, for
a lower density range. For $n \lesssim 1 \times 10^{11}$ cm$^{-2}$, the data points do not decrease linearly as
$n$ is lowered, but appear to extrapolate to a finite field at zero density. To verify that this saturation does
not result from the finite temperature of measurements, we record the $B_{i}$ fields for several temperatures
[Fig.\ 4 (a-c)]. For all measured densities, the fields $B_{min}$ and $B_{mid}$ are approximately independent of
$T$ for $T \lesssim 300$ mK \cite{Bmaxtempdep}. In Fig.\ 4(d) we also plot the temperature dependence of
$R_{xx}(B_{||})$ for the lower two densities in sample A. As for GaAs electrons and holes
\cite{tutuc02,tutuc01,papadakis00,yoon00} and for Si electrons \cite{okamoto99,simonian97}, we observe a
transition from metallic to insulating  behavior near a finite field $B_c$, lower than $B_{min}$. We also note
that the apparent zero-field insulating-to-metallic transition in AlAs 2D electrons occurs in sample B at $n = 0.7
\times 10^{11}$ cm$^{-2}$ \cite{papadakis99}, indicated by a dashed line in Fig.\ 3, which corresponds
approximately to the onset of $B_i$ saturation as $n$ is lowered. A similar saturation for $n < n_c$ seems to also
take place in other 2D systems, such as in GaAs holes \cite{yoon00} and in Si MOSFETs \cite{mertes99}. The physics
behind this phenomenon may be related to a disorder-induced broadening of the spin subbands, but certainly needs
further theoretical and experimental attention.

\begin{figure}
\includegraphics[scale=.45]{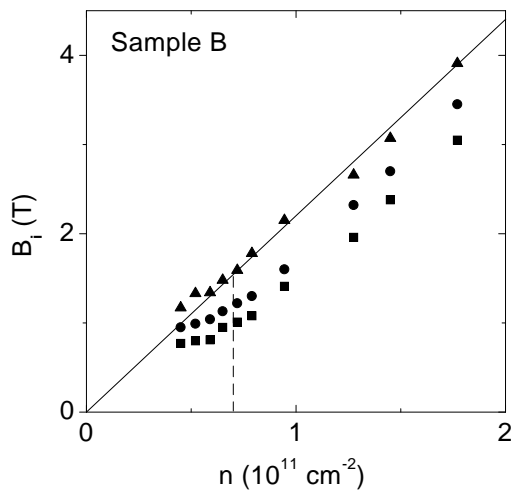}
\caption{$B_{i}$ fields vs.\ $n$ for sample B ($T \simeq$ 300 mK). The fields decrease linearly with decreasing
$n$ over most of the range, but tend to saturate at low $n$. The dashed line indicates the density at the apparent
zero-field metal-insulator transition.}
\end{figure}

\begin{figure*}
\includegraphics[scale=1]{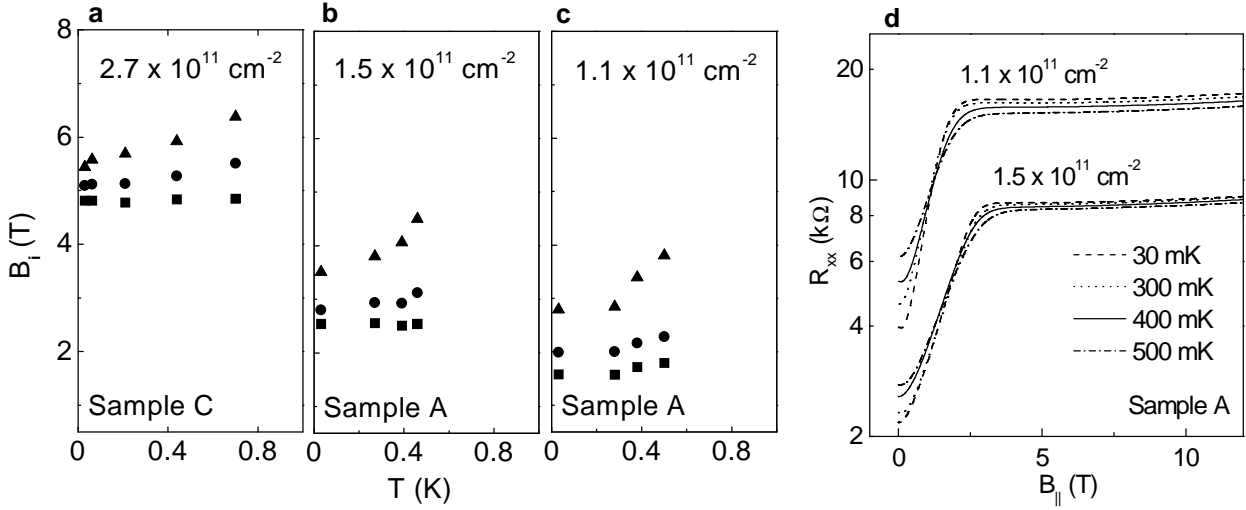}
\caption{(a-c) Temperature dependence of $B_{i}$ fields (defined in text) for three different densities. Data are
shown for sample A in (b) and (c), and for sample C in (a). (d) $T$ dependence of in-plane magnetoresistance of
sample A.}
\end{figure*}

\begin{figure}
\includegraphics[scale=.35]{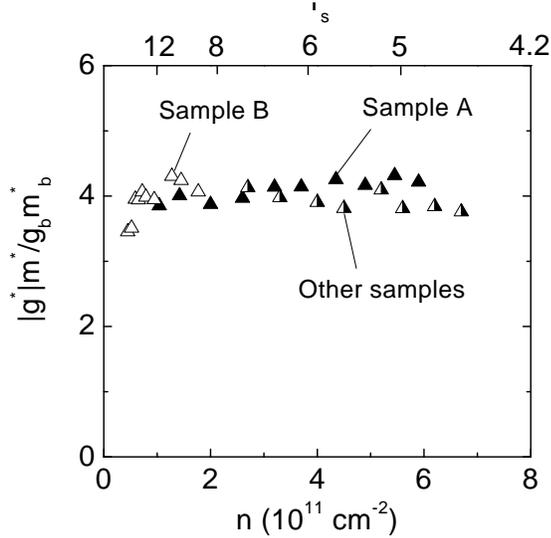}
\caption{Product ($|g^*|m^*$) of the effective $g$-factor and effective mass in AlAs 2D electrons, deduced from
data plotted in Figs.\ 2 and 3, normalized to its band value. $|g^*|m^*$ is remarkably constant over most of the
density range.}
\end{figure}

We conclude this paper by deriving the product $|g^*|m^*$ from our magnetoresistance data, using Eq.\ 1. We assume
that spin subband depopulation occurs at the field $B_{max}$ \cite{tutuc02,tutuc01}, and plot in Fig.\ 5 the
deduced $|g^*|m^*$ for several AlAs samples (including samples A and B \cite{othersamples}), normalized to their
band values using $g_b$ = 2.0, and $m_b^*$ = 0.46 and 0.41 for (100) and (411)B-oriented substrates respectively.
The magnitude of the $|g^*|m^*$ enhancement is approximately equal to 4, a value comparable to the enhanced
$|g^*|m^*$ in GaAs and Si 2D carrier systems for similar $r_s$. Remarkably, we also observe that $|g^*|m^*$ in
AlAs is roughly independent of $n$ for a wide range of densities.

Our results thus contrast with studies of spin polarization in Si MOSFETs, which find that $|g^*|m^*$ increases at
low densities \cite{fang68,okamoto99,pudalov01,shashkin01,vitkalov01}. Authors in Refs.\ 9 and 10 obtain their
conclusions after scaling the magnetoresistance with a single curve independent of $B$ or $T$. We find that for
AlAs 2D electrons, no linear mapping of $B$, $\rho_{xx}$ or $\sigma_{xx}$ ($= 1/\rho_{xx}$) is able to generate
such a scaling law. We thus rely solely on Eq.\ 1 to obtain values for $|g^*|m^*$. Our data therefore do not point
to a ferromagnetic ground state at any finite density, which is consistent with the behavior of GaAs 2D electrons
\cite{tutuc02} and holes \cite{tutuc01,yoon00,papadakis00}.

This work was funded by the NSF and the ARO. We are grateful to Eric Palm and Tim Murphy for their help with
experiments done at the National High Magnetic Field Laboratory, Tallahassee, FL, which is also funded by the NSF.

\break


\begin{references}
\bibitem{valleyoccupation}
In samples A and B, one valley is occupied at $B$ = 0 T for $n \lesssim 3.0 \times 10^{11}$ cm$^{-2}$, while for
larger densities in sample A, two valleys are occupied at low fields. However, we infer from the slope of the
linear fit to $B_{max}(n)$ in Fig.\ 2 that the upper valley depopulates at a field smaller than $B_{min}$, so that
for the purposes of this study, we can assume single valley occupation. In Fig.\ 1, a less pronounced ``kink'' in
$\rho_{xx}$ can indeed be seen at $B \sim 5$ T for the highest density, and may be related to the depopulation of
one of the valleys.
\bibitem{deviation}
For sample B, due to noisier data, $B_{max}$ is defined as the field at which $\rho_{xx}$ reaches 99 $\%$ of its
saturation value.
\bibitem{Bmaxtempdep} Though $B_{max}$ may decrease by $< 10$ \% as $T$ decreases from 300 mK to 30 mK for $n > 1.5 \times
10^{11}$ cm$^{-2}$,  this dependence does not affect our conclusions.
\end{references}
\end{document}